\documentclass[runningheads]{svmult}

\usepackage{makeidx}   
\usepackage{graphicx}  
\usepackage{subeqnar}  
\usepackage{multicol}  
\usepackage{physprbb}  
\makeindex             

\begin{document}
\title*{Variety in Supernovae}
\toctitle{Variety in Supernovae}
%
%
\titlerunning{Variety in Supernovae}
%
\author{Massimo Turatto\inst{1}
\and Stefano Benetti\inst{1}
\and Enrico Cappellaro\inst{2}
}
\authorrunning{Massimo Turatto et al.}
%
%
\institute{INAF, Osservatorio Astronomico di Padova, Vicolo dell'Osservatorio 5,
35122 Padova, Italia
\and INAF, Osservatorio Astronomico di Capodimonte, Via Moiariello 16,
80131 Napoli, Italia}

\maketitle              

\begin{abstract}
Detailed observations of a growing number of supernovae  
have determined a bloom of new peculiar events.
In this paper we take a short tour through the 
SN diversity and discuss some important, physical issues related to it.
Because of the role of SN Ia in determining the cosmological parameters,
it is crucial to understand the physical origin of even subtle, 
observed differences. An important issue is also the reddening correction.
We belive that the measure of interstellar lines on medium resolution 
spectra of SNe can be used to derive lower limits on the 
interstellar extinction.
A few physical parameters of the progenitor, namely radius, mass, density
structure and angular momentum, may explain most of the diversity of 
core-collapse events. In addition, if the ejecta expand into a
dense circumstellar medium the ejecta-CSM interaction
may dominate the observed outcome and provide a mean
to probe the mass loss history of the SN progenitor in the last stages of its
evolution.

\end{abstract}

\section{Introduction}
Despite the natural inclination of human brain to simplify and group phenomena in a
numerable amount of classes, there is no doubt that nature is complex, and its 
complexity increases the more we understand.
This is valid also in astrophysics and, in particular, in the field of Supernovae (SNe).
After the early decades in which the researchers were discovering the universe
of cosmic explosions and phenomenologically grouped them in an handful of
types, we are at a stage in which the detailed analysis shows the 
individuality of each event.
This does not necessarily contradict previous findings, just reflects our
ability to detect, and partially understand, subtler details.
Indeed, if up to the 1980s it was possible to isolate only the two major SN types,
the utilization of linear detectors has lead to a florilegium of types and subtypes.
The present  classification scheme is complex \cite{mt} and accounts, 
in addition to
the intrinsic nature of the explosion mechanisms and progenitors,  
also for
phenomena occurring after the explosion which can dramatically affect the
SN display,
e.g. the interaction of the ejecta with the circumstellar matter. 
In the following we discuss some of the most significant issues related to
the variety of the SNe.

\section{The Diversity of Type Ia SNe}

The thermonuclear explosions of accreting white dwarfs produce type Ia SNe, which
owing to their high luminosity and accurate calibration are successfully used for 
determining the geometry of the Universe \cite{leib00}. This is not to say that
SN Ia are standard candles. In fact, even after excluding a number of extreme 
events or outliers (e.g. SN~1991bg or 1991T), 
spectroscopic and photometric differences among SN Ia do remain.
However, empirical relations between the light curve shape and 
the luminosity at maximum  \cite{phil93,riess98,perl97,phil99} 
allow to recover type Ia SNe as the best distance indicators 
up to cosmological distances.

Where the variance of SN Ia comes from is not yet know even if it probably 
involves differences in the structure and composition of the precursor WD.
Analysis of the very early spectra of SN Ia could give fundamental insights 
in this respect. Unfortunately, because of the very steep rise of luminosity, 
observing SNIa before maximum is a very demanding task.
To address this and other issues related to the physics of type Ia supernovae
a new European collaboration has just started (RTN2-2001-00037) which had a 
cold start with the extensive observations of the early stages of three events, 
SNe~2002bo, 2002dj and 2002er.

\begin{figure}[t]
\begin{center}
\rotatebox{-90}{\includegraphics[width=.8\textwidth]{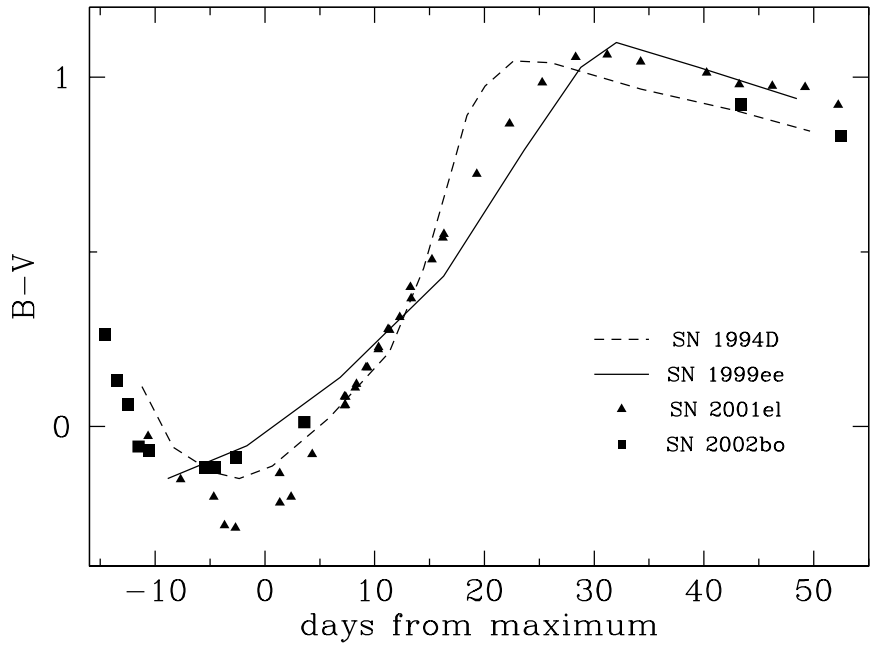}}
\end{center}
\caption[]{Comparison of the preliminary B--V light curve of SN~2002bo with those of other 
normal SNIa, SNe 1994D \cite{pat96}, 1999ee \cite{stritzi} and 2001el \cite{krisci}. 
The color curves have been dereddened according to \cite{phil99}
}
\label{colIa}
\end{figure}

The first data to be analyzed  are those of SN~2002bo. 
The B--V  color curve of SN~2002bo is compared in Fig.~\ref{colIa} with those 
of three other well studied SN Ia after correction for reddening. Indeed, 
following \cite{phil99}
a total reddening E(B-V)=0.48 mag was estimated for SN 2002bo. This is not unexpected
since this SN exploded close to the dust lane of NGC~3190, which obscured  SN~2002cv
by more than 6 magnitudes \cite{meiklecv}.
All objects in the figure show the color evolution typical of SN Ia. 
However, a closer examination of Fig.~\ref{colIa} shows
that significant differences in the  individual behaviors do exist.  
For instance, note that SN~1994D reaches its reddest color about 20-25
days after maximum light, i.e. 10 days before SN~1999ee.
Other differences are found in the premaximum evolution:
while SN~1999ee shows a monotonic reddening starting at least 10 days before maximum,
SN~1994D and more clearly SN~ 2002bo, are very red in the earliest epochs
and reach a minimum (blue) color 3-5days before maximum.

Actually red colors in the early phases of SNIa are predicted by theoretical
models \cite{hoef02,hoef02b}. This is because the decrease
of temperature due to adiabatic cooling  occurs before the heating, due to the 
$\gamma$-rays  from $^{56}Ni$ to $^{56}Co$ radioactive decay, reaches the photosphere.
Since at such epochs the observations sample the outermost layers
of the exploding stars, the differences in the observed color evolution
might reflect differences in the progenitors
structure and composition.

\begin{figure}[t]
\begin{center}
\rotatebox{-90}{\includegraphics[width=.8\textwidth]{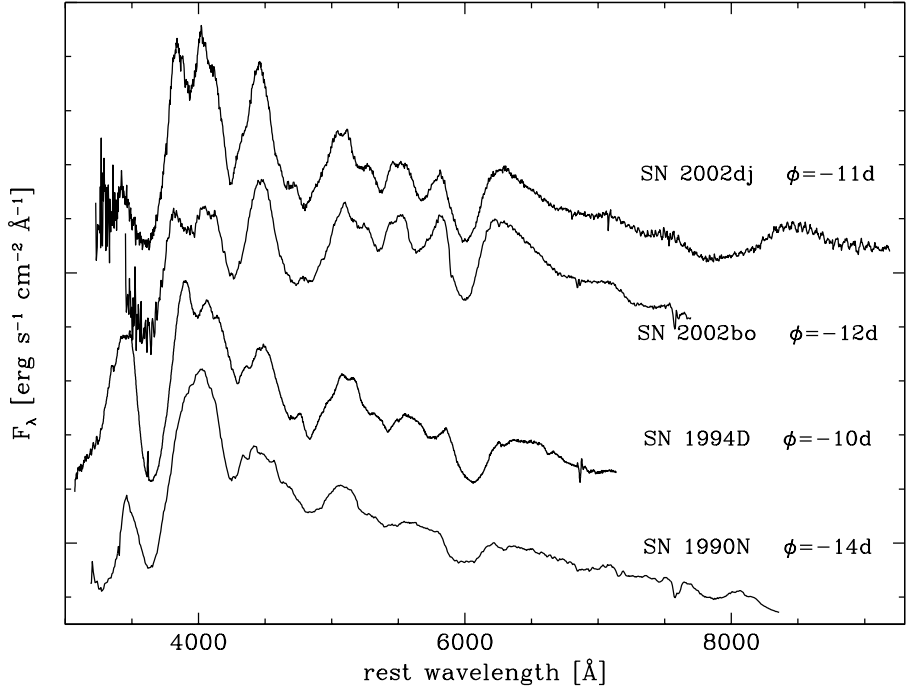}}
\end{center}
\caption[]{Comparison of early--time spectra of SNe 2002bo and 2002dj
with those of 1990N \cite{leib90n} and 1994D (Wheeler, private communication).
The flatter slope of the 
continuum of SN~2002bo is due to reddening
}
\label{spIa}
\end{figure}

Similar considerations can be drawn from the analysis of the spectra.
In Fig.~\ref{spIa} are shown four of the earliest available spectra of 
SNIa (ranging between 14 to 10 days before maximum, that is less than a week
from explosion). Although even small age differences 
may explain part of the observed diversity, thre is no doubt that 
intrinsic differences exist.
In particular, around 6000 \AA\/ we note the different profiles of the SiII 
absorption which in SN~1990N has a flat bottom attributed 
to the contribution of high velocity CII  \cite{fisher}. 
Even more striking is the diversity in the blue side
where entire absorption bands which are visible 
in SN~2002dj, 2002bo and SN~1994D, e.g. that due to SiII 4128, 4131, 
CoII 4161, are absent in SN~1990N. A detailed analysis of the sequence of 
early spectra of SNe 2002bi and 2002dj is in progress.

It should be stressed that the claims of an accelerated expansion of the  
Universe relies  on the comparison of high--z SNe observed in the optical window 
with local templates at blue and UV wavelengths, hence
any uncertainty on the behavior in these bands reflects on the robustness
of the conclusions.

\subsection{The Issue of Reddening}

In general, the light from SNe is absorbed and reddened by interstellar dust 
both in the Galaxy and in the host system. While the Galactic 
component is easily removed using maps of the galactic dust distribution 
\cite{schleg} and a standard extinction 
law, more difficult is evaluating the extinction in the host galaxy. 

A widely adopted method is to estimate the color excess by comparing the SN color
at selected epochs with that of template SNe. In 
particular, it has been suggested that the B$-$V color 2-3 months after 
the explosion is 
independent on the SN photometric class  \cite{phil99}. In addition to require a fairly
good coverage of the SN photometric evolution, this approach relies on the assumption
of an uniform behavior for all SN Ia which should be checked by some independent
measurement.

In principle, high resolution spectroscopy of the interstellar NaID lines
by means of the doublet ratio method can give the gas column density which, 
in turn, can be converted into reddening assuming an average dust-to-gas ratio. 
However, the method has the drawback that, because of the need to reach a good S/N
and a high spectral resolution
it can been applied only to few objects, typically nearby 
SNe observed in proximity of maximum.

An empirical approach has been applied in the past, relating the EW of interstellar
absorption lines, measured on medium resolution SN spectra, 
to the color excess E(B--V) estimated from color curve comparison
\cite{barb89b,rich94}. The first attempts seemed 
to suggest the existence of a simple linear relation which would imply a constant 
dust-to-gas ratio, a unique extinction law and a negligible effect of saturation.
The latter might be understood considering that because of the galaxy rotation 
the various absorbing clouds  have different radial velocity components along the line of
sight. This spreads the interstellar absorptions at different wavelengths and 
prevents heavy line saturation.
However, with the growth of the event statistics the scatter around the relation 
appeared to increase significantly and the existence of a relation was 
questioned. 

\begin{figure}[t]
\begin{center}
\rotatebox{-90}{\includegraphics[width=.8\textwidth]{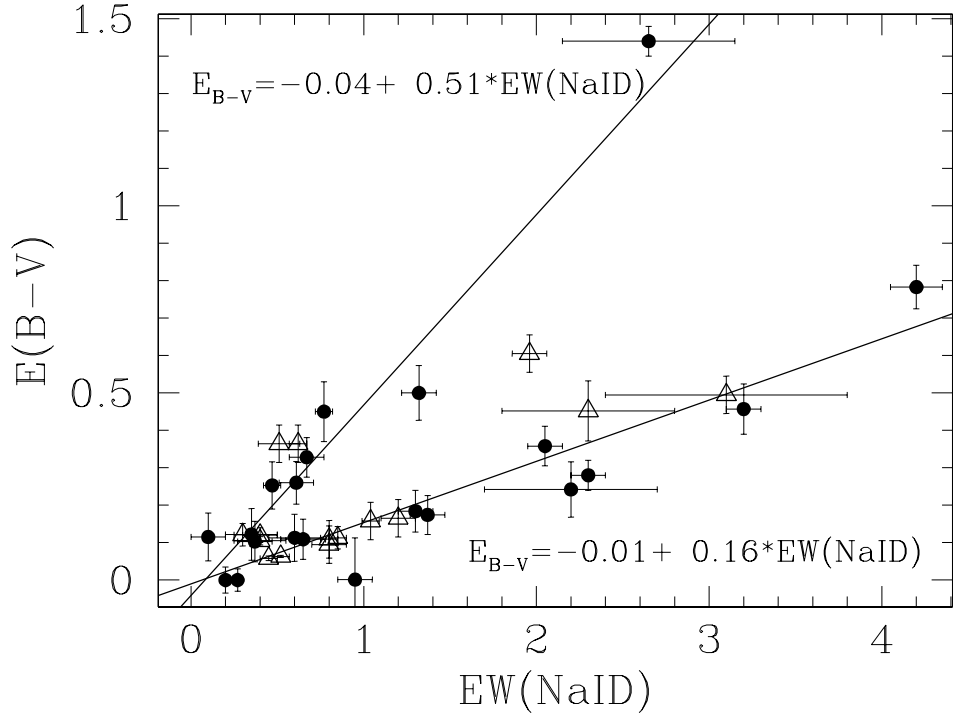}}
\end{center}
\caption[]{Color excess of SNIa inside the host galaxy, 
E(B--V), determined
from the tail of the color curves \cite{phil99}, versus EW(NaID) measured
on low dispersion spectra (filled circles). The values of SN~1986G (the rightmost 
point) include also the Galactic component. 
Open triangles are estimates of Galactic color excess from for SNe
of all types \cite{schleg}}
\label{excessIa}
\end{figure}

To review this issues, we have made use of the homogeneous set of E(B--V) 
estimates for SNIa provided by Phillips et al. \cite{phil99},
integrated by a few more recent objects measured with a similar prescription.
For the SNe of this sample we have searched both in the literature
and in our archive for high signal--to--noise, medium--resolution spectra
and have measured the EW(NaID) of the host galaxy component.
With these data we have redrawn the E(B--V) vs. EW(NaID) plot
(Fig.~\ref{excessIa}).

An accurate examination of the figure shows that the points, although
apparently dispersed, do not fill the plane but rather seem to cluster 
around two lines with significant
different slopes, and only one or two objects in between.
Most objects lie on  the line
with smaller slope ($E(B-V) = 0.16\times EW(NaID)$) which roughly corresponds 
to the previous linear relation \cite{barb89b}. 
However, there are other SNIa which for similar EW(NaID) are much
more heavily reddened ($E(B-V) = -0.04 + 0.51\times EW(NaID)$). 
Interestingly, a similar bivariate 
behavior seems to occur for a sample of SNe of all types also for the
Galactic component, where this time the extinction is derived from 
standard dust maps (open symbols) \cite{schleg}.

The interpretation of this finding is beyond the aim of the present paper and
it is more likely related to  different conditions of the ISM.
Nevertheless it is worth noting that Fig.~\ref{excessIa} tells that
strong reddening is present
each time large values of EW(NaID) are measured in the spectra. 
In other words, by entering in the 
graph with a measurement of the EW(NaID) 
we get lower limits to the reddening of SNIa.

\section{Core--collapse Supernovae}

Several SN types (II, IIb, Ib, Ic, IIn as well as several 
peculiar objects) are thought to explode via the gravitational collapse of the
core.
The great observational diversity has not been fully understood even if it clearly
involves the progenitor masses and configurations at the time of explosion.
Whereas SN IIP are 
thought to originate from isolated massive stars, a generalized scenario has 
been proposed in which common envelope evolution in massive binary systems with varying mass
ratios and separations of the components can lead to various degrees of stripping
of the envelope \cite{nomo95}. According to this scenario the sequence of types
IIL--IIb--Ib--Ic is ordered according to a decreasing mass of the envelope.
\subsection{Energetics}

In the last few years it has become evident that core--collapse SNe can release
different amounts of energy in the explosions. A number of objects with low luminosity
and kinetic energy has been discovered (e.g. \cite{tur97d,zampastrpph}).
Although these events are extensively discussed in other contributions to this workshop
\cite{pastogarch,zampgarch}, we recall here that 
they have probably progenitors with M$\geq20$M$_\odot$. 
In fact, if the progenitor mass is 
large enough core collapse may leave a black hole remnant and 
late time accretion onto the compact remnant can be
a significant source of radiation. Unfortunately the optical 
signature of this event has not been detected yet \cite{ben97d}.

On the other end there are super--energetic SNe, often called \lq hypernovae\lq.
The first and most interesting example was SN~1998bw which 
was discovered while searching for the optical counterpart of GRB980425.
SN ~1998bw was of type Ic, hence believed to originate from 
the core collapse of a
massive star, stripped of its H and He envelope. It was as bright as a SNIa and
the high expansion velocities ($>3 \times 10^4$ km s$^{-1}$ )
indicate that it was unusually energetic ($> 10^{52}$ ergs) \cite{pat98bw}. 
Its very powerful radio emission has been
attributed to a mildly relativistic blast wave interacting
with a clumpy, stratified CSM deriving from a turbulent mass--loss history 
\cite{weil98bw}. Other SNe (e.g. SN~1997ef, SN~1998ey and 
SN~2002ap) bear some spectroscopic resemblance to SN~1998bw 
but are slightly less energetic. 

In all these cases the masses of the progenitors estimated by fitting
light curves and spectra are larger than in normal core collapse 
SNe \cite{nomo02}.  In a qualitative scenario which may explain this finding,
the outcome  of the core collapse of stars with M$\geq 20-25$ M$_\odot$ 
results in under-- or hyper--energetic explosions, depending on 
the angular momentum of the collapsing cores. Faint, 
slowly expanding SNe like 1997D occur because 
the progenitor envelopes have a large binding 
energy and relatively little energy remains available for heating up and 
accelerating
the ejecta. Instead if the core of the progenitor is in rapid rotation, 
owing possibly to the spiraling--in of a companion star, a 
high energetic, asymmetric explosions may be obtained \cite{macf}.

An intriguing possibility is that SN~1997cy, 
observationally classified as SNIIn and possibly
the brightest SN ever observed \cite{germ99,tur97cy}, and its twin SN~ 1999E 
\cite{rigon}, are associated to GRBs. 
As in the case of other SNIIn, these events show strong ejecta-CSM interaction
with explosion energies as high as $3 \times 10^{52}$ ergs.

\begin{figure}[t]
\begin{center}
\rotatebox{-90}{\includegraphics[width=.8\textwidth]{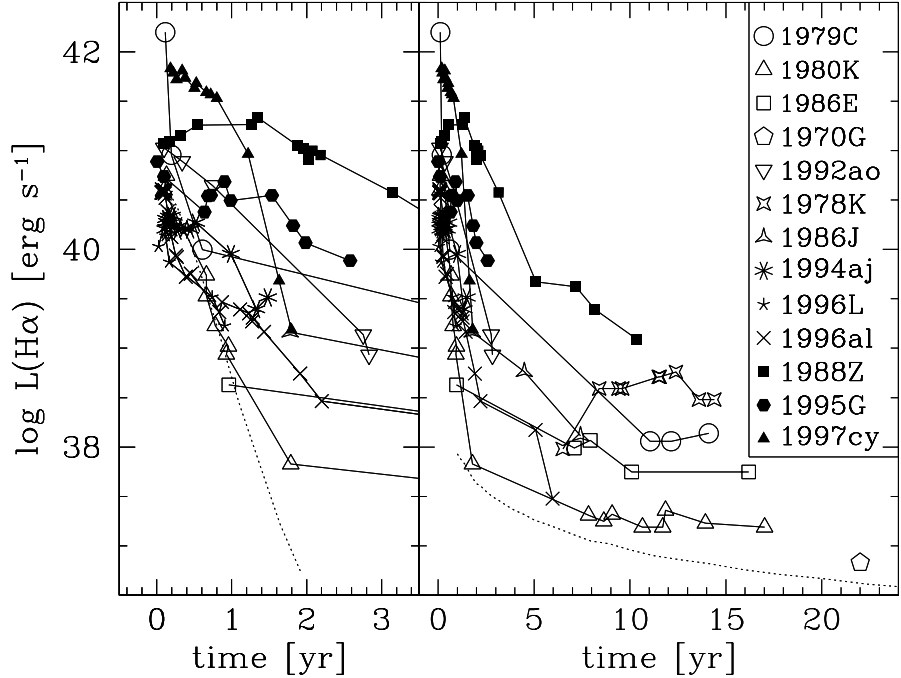}}
\end{center}
\caption[]{The long--term H$\alpha$ evolution of interacting SNII.
On the left panel is shown as a dotted curve the radioactive 
model \cite{chug} while on the right-hand panel is the interaction 
model for SN~1980K \cite{cf}.
SNe not following the radioactive model require an additional source of energy, which
is provided by the interaction with the CSM.  SNIIn (filled symbols)
decline slowly starting soon after explosion 
\cite{tur93,aret,tur97cy,pas02}. Some SNII with linear light curves
\cite{fesen,capp95,ben94aj,ben96l,fesen93} (open symbols) 
show evidence of interaction at late stages.   
Also the optical observations of the bright radio 
SNe 1986J and 1978K are reported \cite{chu95,leib91}
Unpublished data of SN~1992ao, 1996al (ESO 3.6m and NTT) and 1986E (ESO-VLT)
come from the Padova archive.}
\label{halphaflux}
\end{figure}

\subsection{Interaction with the CSM}

Very important in determining the outcome of core--collapse SNe is 
the possible presence of a dense CSM around the progenitor star.
Indeed the interaction of the fast ejecta with the slowly expanding CSM
generates a forward shock in the CSM and a reverse shock in the ejecta. The shocked
material emits radiation in the optical, radio and X--rays with 
characteristics which depends on the density of both the CSM and the ejecta, 
and on the properties of the shock \cite{chevfr94}. Studying the ejecta--CSM 
interaction the mass loss history 
in the late stages of the stellar evolution can probed.

In some cases it turns out that the interaction begins immediately after
the burst indicating that strong wind persisted
up to the very last stages of 
progenitor life. Often the radiation from the shocked region shadows
the thermal emission from the ejecta and the spectrum is dominated by 
strong emission lines with composite profiles, which reflects the different 
kinematics of the emitting layers. Observationally these SNe are called type IIn.

In most cases, the SN initially expands in an empty space and no interaction occurs.
However, in a number objects the fast expanding ejecta eventually
catches the material ejected during remote strong wind events  
and the SN emission is revived by the interaction.
With the noticeable exception SN~1957D, an unclassified, poorly studied object,
the optical spectra of late--time revived SNe are
dominated by  broad, boxy H emission \cite{fesen,capp95}. 

The different behaviors of the flux evolution of H$\alpha$ for a number of CSM interacting
SNII is displayed in Fig.~\ref{halphaflux}.
In general, type IIn SNe show a slow evolution from the earliest
phases and the H$\alpha$ can
remain almost constant for years, like in the cases 
SN~1988Z and SN~1995G \cite{aret,pas02}.

The onset of the interaction at late times  is evident in the well studied SNe 
1979C, 1980K and 1986E \cite{fesen,capp95}. 
After a rather normal evolution in which the
H$\alpha$ flux matches the radioactive decay input energy \cite{chug} 
(dotted curve in the left panel Fig.~\ref{halphaflux}),
months or years after the explosion the radiation from interaction
becomes dominant, the line flux almost halts its decline and the temporal evolution
is well reproduced by
the circumstellar interaction model (dotted curve in the right panel) \cite{cf}.
These three SNe belong to the subclass with linear light curves (SNIIL) which
are thought to have 
lost most their hydrogen envelope before the explosion.
Therefore, the delayed onset of
ejecta--CSM interaction is not surprising.  In addition, all of them have been 
detected in the radio, supporting the claim that late time
optical and radio emission are correlated \cite{capp95}.
Similar behaviors, i.e. linear light curves, flattening of the H$\alpha$ flux
evolution and boxy line profiles at late time have been exhibited also by SNe 1994aj, 
1996L and 1996al, which showed signatures of slowly expanding shells
above the photosphere also in the earliest times \cite{ben94aj,ben96l}.

Although differences between the observations and the models
do exist and improved interaction models are needed, we emphasize
that the present observational facilities make possible to
study in detail the interaction of the ejecta with the CSM 
(kinematics, density, temperature the emitting layers) for decades 
after the explosion which 
corresponds to probe tens of thousands of years of the progenitor mass loss history.

\section{Conclusions}

In this paper we have addressed the SN variety.
SNIa, which are used as distance indicators up to cosmological
distances, can differ as to absolute magnitudes, intrinsic colors at maximum, light
and color curve shapes. These differences might be related to the composition and
structure of the WD and to variations in the explosion mechanism.

All SNe can be heavily reddened. In the plane EW(NaID) vs. E(B-V), type Ia SNe seem 
to cluster on two different 
linear relations, possibly due to different dust-to-gas
ratio in the host galaxies. If, on one side this prevents that 
accurate reddening can be obtained only by means of medium dispersion
spectroscopy, on the other there is no doubt  
that significant reddening is always present when interstellar absorption
lines are observed (see Fig.~\ref{excessIa}). 

Core--collapse SNe show a wider variety. Although all ignited by the same 
event, i.e. the core-collapse, the explosion 
of progenitors with  radii, masses, density structures and angular momenta
different up to one order of magnitude can  release different amounts of energy
and  variable amounts of heavy and intermediate mass elements. These, in turn,
result in significantly different observables (absolute magnitudes, colors,
spectral and luminosity evolution, etc.).

Despite the total number of SNe discovered is well over 2300 \cite{barb99}, 
there is no doubt that we are still scratching the surface 
of the SN diversity.


%

\end{document}